\documentclass[preprintnumbers,amssymb]{revtex4}
\usepackage{amssymb}
\usepackage{amsmath}
\usepackage{graphicx}
\usepackage{subfig}
\usepackage{dcolumn}
\usepackage{bm}

\begin{document}
\addtolength{\voffset}{-.75cm}
\addtolength{\textheight}{1.7cm}
\addtolength{\hoffset}{-0.25cm}
\addtolength{\textwidth}{.55cm}
\title{Apparent CPT Violation in Neutrino Oscillation Experiments}
\author{Netta Engelhardt$^{(1,2)}$, Ann E. Nelson$^{2}$, and Jonathan R. Walsh$^2$}
\affiliation{\\
$^1$Department of Physics, Brandeis University, 415 South St., Waltham, MA 02454\\ 
$^2$Department of Physics, Box 1560, University of Washington,
           Seattle, WA 98195
}
\newcommand{\geff}{g_{\rm eff}}
\newcommand{\meff}{m_{\rm eff}}
\newcommand{\p}[0]{\partial}
\renewcommand{\d}[0]{\textrm{ d}}
\newcommand{\R}[0]{\mathbb{R}}
\newcommand{\C}[0]{\mathbb{C}}
\newcommand{\bra}[1]{\big<#1\big|}
\newcommand{\ket}[1]{\big|#1\big>}
\renewcommand{\matrix}[4]{\left( \begin{array}{c c} #1 & #2 \\ #3 & #4 \end{array} \right)}
\newcommand{\pf}[2]{\frac{\partial #1}{\partial #2}}
\newcommand{\df}[2]{\frac{\textrm{d}#1}{\textrm{d}#2}}
\newcommand{\be}[0]{\begin{equation*}}
\newcommand{\ee}[0]{\end{equation*}}
\newcommand{\sfrac}[2]{\textstyle{\frac{#1}{#2}}}
\newcommand{\Res}[0]{\textrm{Res}}
\newcommand{\lc}[1]{\epsilon_{#1}}
\newcommand{\del}[0]{\nabla}
\newcommand{\braket}[2]{\big<#1\big|#2\big>}
\newcommand{\beq}{\begin{eqnarray}}
\newcommand{\eeq}{\end{eqnarray}}
\newcommand{\nn}{\nonumber}
\newcommand{\bml}{ $U(1)_{\textrm{\tiny{B-L}}}$}
\def\ltap{\ \raise.3ex\hbox{$<$\kern-.75em\lower1ex\hbox{$\sim$}}\ }
\def\gtap{\ \raise.3ex\hbox{$>$\kern-.75em\lower1ex\hbox{$\sim$}}\ }
\def\CO{{\cal O}}
\def\CL{{\cal L}}
\def\CM{{\cal M}}
\def\tr{{\rm\ Tr}}
\def\CO{{\cal O}}
\def\CL{{\cal L}}
\def\CM{{\cal M}}
\def\mpl{M_{\rm Pl}}
\newcommand{\Dslash}{D\llap{/\kern+2.1pt}}
\newcommand{\bel}[1]{\be\label{#1}}
\def\al{\alpha}
\def\bt{\beta}
\def\eps{\epsilon}
\def\eg{{\it e.g.}}
\def\ie{{\it i.e.}}
\def\mn{{\mu\nu}}
\newcommand{\rep}[1]{{\bf #1}}
\def\be{\begin{equation}}
\def\ee{\end{equation}}
\def\bea{\begin{eqnarray}}
\def\eea{\end{eqnarray}}
\newcommand{\eref}[1]{(\ref{#1})}
\newcommand{\Eref}[1]{Eq.~(\ref{#1})}
\newcommand{\gsim}{ \mathop{}_{\textstyle \sim}^{\textstyle >} }
\newcommand{\lsim}{ \mathop{}_{\textstyle \sim}^{\textstyle <} }
\newcommand{\vev}[1]{ \left\langle {#1} \right\rangle }
\newcommand{\ev}{{\rm eV}}
\newcommand{\kev}{{\rm keV}}
\newcommand{\Mev}{{\rm MeV}}
\newcommand{\gev}{{\rm GeV}}
\newcommand{\tev}{{\rm TeV}}
\newcommand{\mev}{{\rm MeV}}
\newcommand{\mnu}{\ensuremath{m_\nu}}
\newcommand{\mlr}{\ensuremath{m_{lr}}}
\newcommand{\acc}{\ensuremath{{\cal A}}}
\newcommand{\mav}{MaVaNs}
\newcommand{\nusm}{$\nu$SM }

\begin{abstract}
We consider searching for  light sterile fermions and   new forces by using long baseline oscillations of neutrinos and antineutrinos.  A new light sterile state and/or a new force  can lead to apparent CPT violation in muon neutrino and antineutrino oscillations. As an example, we present an economical model of neutrino masses containing a sterile neutrino. The potential from the Standard Model weak neutral current gives rise to a difference between the disappearance probabilities of  neutrinos and antineutrinos, when mixing with a light sterile neutrino is considered.  The addition of a  $B-L$ interaction adds coherently to the neutrino current potential  and increases the difference between neutrino and antineutrino disappearance.  We find that this model can improve the fit to the results of MINOS for both neutrinos and antineutrinos, without any CPT violation, and that the  regions of parameter space which improve the fit  are within experimental  constraints.  
\end{abstract}
\maketitle
\section{Introduction}
Non-collider experiments and astronomical observations  have given us our first hints of physics beyond the Standard Model, via the discoveries of neutrino oscillations, dark energy, and dark matter. The implication of these discoveries for fundamental physics is still unknown. The energy scale of the new degrees of freedom giving rise to neutrino oscillations could be as high as $10^{16}$ GeV, as in Grand Unified theories, or as low as $0.05$ eV, as in Dirac neutrino mass models. Even more mysterious is the nature of  dark energy and dark matter, and the  associated energy scale or scales. If the new physics is light, it must be very weakly coupled to the Standard Model, or it would already have been discovered. Neutrino oscillation measurements offer an unmatched portal into any new nonstandard sectors containing light fermions, because neutrinos   can mix with neutral spin 1/2 particles, and because oscillations over long baselines are extraordinarily sensitive to extremely tiny effects. 

The long baseline experiment MINOS, which uses similar   near and far detectors to reduce systematic errors,   has observed the disappearance of both  muon neutrinos and muon antineutrinos in the far detector \cite{Michael:2006rx,Gogos:2007zz,Adamson:2008zt,Evans:2008zzc}. The antineutrino data comes from a 7\% antineutrino contamination of the beam and is severely statistics limited \cite{Adamson:2009ju}. Nonetheless, it is interesting to note that
the antineutrino disappearance rate is larger than the rate expected from  neutrino disappearance    by almost 2 sigma \cite{Evans:2008zzc}. Currently MINOS is running in antineutrino mode, offering a unique opportunity to precisely measure the parameters governing the long baseline oscillations of   muon antineutrinos. It is the purpose of this paper to offer a framework for searching for new physics in the antineutrino data. Most recent analyses \cite{Evans:2008zzc,Dighe:2008bu,Koranga:2009ep,Arias:2009fk,Barenboim:2009ts,Diaz:2009qk,Arias:2009fk} of anomalies in antineutrino data have focussed on CPT violation in the neutrino mass matrix
 \cite{Murayama:2000hm,Barenboim:2001ac,Skadhauge:2001kk,Bilenky:2001ka,Jacobson:2003wc,Kostelecky:2003cr}.  However there is no theoretical motivation for CPT violation, and CPT violation requires Lorentz violation which is complicated to incorporate in a complete theory that is consistent with other data. In contrast, in this paper we propose a simple,   renormalizable, Lorentz invariant field theory, which is  consistent with other experiments and which allows  a significant difference  between muon neutrino and antineutrino disappearance in the MINOS experiment.  
 
Many other papers have analyzed the consistency of neutrino oscillation data with sterile neutrinos \cite{Barger:1990bg,Goswami:1995yq,GonzalezGarcia:1998dp,Maltoni:2004ei,GonzalezGarcia:2007ib,Maltoni:2007zf} and new forces \cite{Valle:1987gv}. Our model differs from those considered previously in three ways.  First, we consider a relatively long range but weakly coupled new force for which the size of the matter effect can be considerably larger than the usual MSW effect, in a theory which is nevertheless consistent with precision electroweak constraints.  Second, we   consider smaller neutrino mass squared differences that primarily affect long baseline experiments for muon neutrinos. Thirdly, many previous analyses were concerned with the effects of sterile neutrinos on electron neutrino oscillations, while we are not attempting to address any anomalies involving electron neutrinos and are considering a model where the sterile neutrino has no electron neutrino component.

\section{Review of the standard picture of neutrino oscillations}
A standard picture  of 3 flavor neutrino oscillations has been  successful in explaining  phenomena observed by many diverse long baseline experiments \cite{Maki:1962mu,GonzalezGarcia:2010er}.  
The $e$, $\mu$, and $\tau$ flavor eigenstates are related to the mass eigenstates via a unitary transformation,  parameterized by three angles, neglecting a CP-violating phase:
\be\label{stdosc}
\left( \begin{array}{c}
\nu_e \\
\nu_{\mu} \\
\nu_{\tau}
\end{array} \right) = \left( \begin{array}{c c c}
1 & 0 & 0 \\
0 & c_{23} & s_{23} \\
0 & -s_{23} & c_{23}
\end{array} \right)\left( \begin{array}{c c c}
c_{13} & 0 & s_{13} \\
0 & 1 & 0 \\
-s_{13} & 0 & c_{13}
\end{array} \right)\left( \begin{array}{c c c}
c_{12} & s_{12} & 0 \\
-s_{12} & c_{12} & 0 \\
0 & 0 & 1
\end{array} \right)\left( \begin{array}{c}
\nu_1 \\
\nu_2 \\
\nu_3
\end{array} \right)\equiv U_{LNS}\left( \begin{array}{c}
\nu_1 \\
\nu_2 \\
\nu_3
\end{array} \right)\ ,
\ee
where $c_{ij} = \cos\theta_{ij}$, $s_{ij} = \sin\theta_{ij}$.  A good fit to long baseline neutrino oscillation data may be found \cite{Amsler:2008zzb} for angles
 \be\label{stdangles}
\begin{split}
\sin^2 2 \theta_{12} &= 0.87\pm0.03,\\
\sin^2 2\theta_{13} &< 0.19 \textrm{ (90\% C.L.}),\\
\sin^2 2\theta_{23} &= 1.0^{+0}_{-0.08} ,
\end{split}
\ee
and  neutrino mass squared differences:
\be\label{stdmasses}
\begin{split}
\Delta m_{12}^2 &= (7.6\pm0.2)\cdot10^{-5}\textrm{ eV}^2/c^4 ,\\
\big|\Delta m_{23}^2\big| &= (2.4\pm0.1)\cdot10^{-3}\textrm{ eV}^2/c^4 .
\end{split}
\ee
Note that in this picture,  for $L/E < 1000$ km/GeV,  the smaller mass squared difference $\Delta m_{12}^2$ gives oscillation probabilities which are always less than a percent, and which may be neglected compared with experimental uncertainties. Therefore for $L/E < 1000$ km/GeV,  the observed oscillations  are only sensitive to  $\Delta m_{23}^2$ and, due to the small size of $\theta_{13}$, are primarily    muon flavor into tau flavor.

In matter, the oscillation parameters of  neutrinos are modified due to the weak potential, a phenomena known as the MSW effect \cite{Wolfenstein:1977ue,Mikheev:1986gs,Langacker:1986jv}. In the standard picture, only oscillations involving $\nu_e$ from a charged current weak potential are affected, as the $\nu_\mu$ and $\nu_\tau $ neutrinos only experience a flavor universal, neutral current weak potential in matter.  

The  MINOS experiment  studies the oscillations of  both  muon neutrinos and muon antineutrinos at a distance of 735 km, over an energy range from 1 to 50 GeV. The survival probability as a function of energy for muon neutrinos, $P(\nu_{\mu}\to\nu_{\mu})$, is well parameterized by Eqs.~\ref{stdangles} and \ref{stdmasses}.  The survival probability for muon antineutrinos, $P(\bar{\nu}_{\mu}\to\bar{\nu}_{\mu})$ is only marginally consistent with the standard picture,  and is not well characterized by the oscillation parameters in Eqs.~\ref{stdangles} and \ref{stdmasses}.  The marginal agreement of  the neutrino and antineutrino oscillation data at MINOS has  raised interest,  particularly because in vacuum, if CPT is a good symmetry, then the two oscillation probabilities are equal:
\be
\textrm{CPT} \quad \Rightarrow \quad P(\nu_{\mu}\to\nu_{\mu}) = P(\bar{\nu}_{\mu}\to\bar{\nu}_{\mu})\, ,
\ee
and in the standard oscillation picture, the presence of matter does not significantly change this relation at the MINOS baseline.

\section{Sterile neutrinos, a new force, and MINOS }
The MINOS antineutrino disappearance data has led several authors to investigate the possibility that  CPT violation may have been observed at MINOS.   However, because the neutrinos are passing through matter,  alternative explanations include a nonstandard  interaction between the neutrinos and the matter  which distinguishes between neutrinos and antineutrinos,  a new kind of neutrino with a different weak charge, or both.

In this work we discuss a model that can generate  a difference between muon neutrino and muon antineutrino disappearance probabilities in matter at the MINOS  energies and baseline.  The basic framework is   simple. We extend the Standard Model to allow the active neutrinos to mix with a sterile neutrino that has no Standard Model weak interaction. Because a sterile neutrino has a different weak charge than the active neutrinos, in matter   muon neutrinos and muon antineutrinos will have different oscillation parameters, although this effect is small and  only marginally detectable at the MINOS baseline.  

In Ref.~\cite{Nelson:2007yq}, two of us 
proposed a new $U(1)$ gauge force, coupled to the difference between baryon and lepton numbers ($B-L$), which can enhance this matter effect, in order to account for the anomalous results of the MiniBooNE and LSND experiments. We showed that provided the new boson was very  weakly coupled ($g\lsim 10^{-5}$),  such a boson of mass greater than $\approx0.2 $~keV could escape all laboratory constraints.  If such a weakly coupled boson is  lighter than a few  MeV, it can provide a potential in the earth's crust which is much larger than the standard MSW potential. Furthermore, in Ref.~\cite{Nelson:2008tn} we showed that  a light vector boson could have a chameleon nature, with very different effective mass in extreme environments, allowing it to escape astrophysical  and cosmological constraints. 
Due to the very different energies and baselines of LSND, MiniBooNE, and MINOS,   a minimal  model which gives an observable anomalous  neutrino/antineutrino oscillation difference at MINOS would not necessarily  significantly affect LSND and MiniBooNE.  

The best fit points that we will find for the model from the MINOS neutrino and antineutrino oscillation data have  a contribution to the potential from the $B-L$ interaction which is approximately 4 orders of magnitude smaller than the one required to  reconcile  the shorter baseline MiniBooNE and LSND experiments.  This allows the $B-L$ coupling $g$ to be lower by 2 orders of magnitude for the same $B-L$ vector boson mass.  Consequently, the constraints on the $B-L$ force discussed in Ref.~\cite{Nelson:2007yq} do not limit the model in describing neutrino/antineutrino oscillation anomalies at MINOS, and in general at long baseline oscillation experiments.

As in the usual seesaw model of neutrino masses \cite{GellMann:1976pg,Minkowski:1977sc,Weinberg:1979sa,Mohapatra:1979ia,Bilenky:1980cx,Yanagida:1980xy,Lazarides:1980nt,Weinberg:1980bf,Schechter:1980gr,Cheng:1980qt}, we consider two types of neutrinos -- neutrinos $\nu_i$ with lepton number $L=1$ which carry Standard Model weak interactions  and neutrinos  $N_i$ with $L=-1$ which are sterile under the Standard Model. In general the number of sterile neutrinos is not fixed by terrestrial experiment.   We allow for both lepton number conserving and violating mass terms, with both types of mass terms   required to give active-sterile mixing.  The allowed  lepton number conserving mass terms, which arise from a Yukawa coupling to the Standard Model Higgs, are
\be
m_{ij}\nu_iN_j + \textrm{ h.c.}\ .
\ee
We also have  lepton number violating Majorana masses for the sterile neutrinos   
\be
M_{ij}N_iN_j + \textrm{ h.c.}
\ee
The Majorana terms are gauge invariant under the Standard Model. With gauged $B-L$, they may  arise from an expectation value of  a $B-L$ charged scalar. 
Unlike in the usual seesaw model, we assume that at least one of the sterile neutrinos is very light. For simplicity of analysis we will  assume any other sterile neutrinos, such as would be required to cancel a gauge $B-L$ anomaly,  are heavier and more weakly mixed and will integrate them out. We thus consider an oscillation  picture involving  4 light neutrinos-- three active   and one   sterile.

Neutrinos are produced and detected as one of the three active neutrino flavors, which propagate as a superposition of all 4 mass  eigenstates.  A 4-by-4 unitary matrix  $U$ transforms between the flavor and mass eigenstate bases. In general,  four flavor neutrino oscillations are affected by 6 mixing angles and three $CP$ violating phases.  We   make some   assumptions to reduce the number of free parameters and simplify the analysis. We  assume the neutrino masses to be hierarchical, with the heaviest mass eigenstate mostly sterile. We assume the sterile neutrino  is a mixture of the 2 heaviest states only.   We can then write the mixing matrix as \be U=U_aU_s,\ee where
$U_a$  is  a block diagonal matrix 
\be U_a=\left( \begin{array}{c c c c}
&   & & \\
  & U_{LNS} &   &\\
  &   & &1
\end{array} \right)\ee
representing mixing between the three active neutrinos and 
\be U_s=\left( \begin{array}{c c c c}
1& 0&0&0 \\
0 & 1&0  &0\\
0  & 0 &\cos\theta_{34} &\sin\theta_{34} \\
0&0&-\sin\theta_{34} &\cos\theta_{34} 
\end{array} \right)\ee  
describes mixing between active and sterile neutrinos.  In the flavor basis, the Hamiltonian is
\be
\mathcal{H}_f = \frac{1}{2E}U_aU_s \mathcal{M}_m^2 U_s^T U_a^T + \mathcal{V}_f .
\ee
$\mathcal{M}_m$ is the mass matrix in the mass basis and $\mathcal{V}_f$ is the potential matrix in the flavor basis,
\be
\mathcal{V}_f = {\rm diag}(V_{cc} - V_{nc} - V_{B-L},\,-V_{nc} - V_{B-L},\,-V_{nc} - V_{B-L},\, V_{B-L}) .
\ee

The potential $V_{cc}$ arises from the weak charged current,  $V_{nc}$ from the   weak neutral current, and $V_{B-L}$ from a new $B-L$ gauge interaction. The values are:
\be
V_{cc}=\pm\sqrt2 G_F N_e, \textrm{   } V_{nc} = \pm \frac{G_F }{\sqrt2}N_n, \textrm{ and } V_{B-L} = \frac{\tilde{g}^2}{m_V^2}N_n\, ,
\ee
where $N_e$ is the electron density, $N_n$ is the neutron density, $\tilde{g}$ is the $B-L$ coupling constant, and $m_V$ is the mass of the $B-L$ vector boson.   With a reasonable matter density of $\rho \approx 2.7$ g/cm$^3$, the neutral current potential has magnitude
\be
V_{nc} \approx 5.0\cdot10^{-5}\textrm{ neV}.
\ee
We will define the parameter $V \equiv \frac12 V_{nc} + V_{B-L}$, which will arise in the oscillation probability formula.  The factor of $1/2$ arises because the neutral current potential only affects active neutrinos. $V$ in matter is positive for neutrinos and negative for antineutrinos. 

For simplicity, and because the MINOS range of  $L/E$ is not large enough to be sensitive to the smaller mass squared differences, we neglect the 2 smaller masses and take
 \be
\mathcal{M}_m = {\rm diag}(0,0,m,M) .
\ee
With this restricted mixing, $U_{12}$ commutes with $U_s$ and $\mathcal{M}_m$.  Therefore, we can neglect $U_{12}$ in $U_a$, since it does not contribute to oscillations.  For simplicity, we also neglect the small angle   $\theta_{13}$.  Then the electron neutrino does not significantly participate in oscillations at the MINOS baseline, meaning $V_{cc}$ is irrelevant and $U_a$ commutes with $\mathcal{V}_f$.     We can further reduce parameters by assuming   the mass $m$ results entirely from mixing with the light sterile neutrino via the seesaw mechanism, receiving no significant contribution from the heavier neutrinos we have integrated out.  The active-sterile  mixing angle   then satisfies
\be
\tan \left(2\theta_{34}\right) = \frac{2\sqrt{Mm}}{M-m} .
\ee
We can now write the Hamiltonian    in an interaction basis where .
\be
H_f = U_aU_s' H_i U_s'{}^TU_a^T ,
\ee
with
\be
U_s'H_i U_s'{}^T = \frac{1}{2E}U_s\mathcal{M}_m^2U_s^T + \mathcal{V}_f .
\ee
The diagonalized Hamiltonian is
\be
H_i = \frac{1}{2E}{\rm diag}\left(0,0,\tilde{m}^2,\tilde{M}^2\right) ,
\ee
and the   mixing matrix  between the sterile neutrino and a linear combination of $\nu_\mu$ and $\nu_\tau$ is
\be
U_s' =  \left( \begin{array}{c c c c}
1 & 0 & 0 & 0 \\
0 & 1 & 0 & 0 \\
0 & 0 & \cos\theta_s & \sin\theta_s \\
0 & 0 & -\sin\theta_s & \cos\theta_s 
\end{array} \right) .
\ee
In terms of the other parameters,
\be\label{mangle}
\begin{split}
\tan\left( 2\theta_s\right) = \frac{\sin\left( 2\theta_{34}\right)}{\cos\left(2\theta_{34}\right) + \alpha} ,
\end{split}
\ee
where
\be
\alpha \equiv \frac{4EV}{M^2 - m^2} .
\ee
Note that this mixing angle is larger in the antineutrino sector,  where $\alpha$ is negative, and is maximal   for
\be\cos(2\theta_{34}) = \frac{-4EV}{M^2 - m^2} \quad \Rightarrow \quad |4EV| = (M-m)^2 \ . \ee 
This effect can give substantial conversion of muon antineutrinos into sterile fermions  for a range of energies at the MINOS baseline.
The eigenvalues of the Hamiltonian are
\be\label{eigen}
\begin{split}
\tilde{m}^2 &= m^2 + \frac{M^2 - m^2}{2}\left[1 + \alpha - \sqrt{1 + 2\alpha\cos(2\theta_{34}) + \alpha^2}\right] ,\\
\tilde{M}^2 &= m^2 + \frac{M^2 - m^2}{2}\left[1 + \alpha + \sqrt{1 + 2\alpha\cos(2\theta_{34}) + \alpha^2}\right] .
\end{split}
\ee
The unitary matrix relating the interaction and flavor bases is
\be
U_aU_s' =  \left( \begin{array}{c c c c}
1 & 0 & 0 & 0\\
0 & c_{23} & s_{23}c_s & s_{23}s_s \\
0 & -s_{23} & c_{23}c_s & c_{23}s_s \\
0 & 0 & -s_s & c_s 
\end{array} \right) ,
\ee
with $c_s = \cos\theta_s$, $s_s = \sin\theta_s$.  Taking  $\theta_{23} = \pi/4$ gives the simple mixing matrix
\be
U_aU_s' =  \left( \begin{array}{c c c c}
\vspace{0.05in}
1 & 0 & 0 & 0 \\
\vspace{0.05in}
0 & \frac{1}{\sqrt2} & \frac{1}{\sqrt2}c_s & \frac{1}{\sqrt2}s_s \\
\vspace{0.05in}
0 & -\frac{1}{\sqrt2} & \frac{1}{\sqrt2}c_s & \frac{1}{\sqrt2}s_s \\
0 & 0 & -s_s & c_s 
\end{array} \right) .
\ee
The muon  neutrino  survival probability is then
\be
P(\nu_\mu\to\nu_\mu) = 1 - \cos^2\theta_s \sin^2\left(\frac{\tilde{m}^2 L}{4E}\right) - \sin^2\theta_s \sin^2\left(\frac{\tilde{M}^2 L}{4E}\right) - \cos^2\theta_s\sin^2\theta_s \sin^2\left(\frac{(\tilde{M}^2 - \tilde{m}^2)L}{4E}\right) .
\ee
The same formula applies to antineutrinos, although the values of $\tilde m, \tilde M$ and $\theta_s$ are different because of the opposite sign of $\alpha$ in Eqs.~\ref{mangle} and \ref{eigen}.

\section{Comparison of Neutrino Oscillation Model with MINOS Data}
Armed with the muon neutrino survival probability, we can constrain the parameter space of the model using the results of the MINOS experiment.  We will perform a combined fit of the model parameters $m$, $V$, and $M$ using the MINOS $\nu_{\mu}$ and $\bar{\nu}_{\mu}$ survival data.  We briefly discuss implications for the K2K \cite{Ahn:2002up} and SuperK \cite{Fukuda:1998mi} experiments, and the  constraints from the CDHS results \cite{Dydak:1983zq}. 

We perform a $\chi^2$ fit to the MINOS $\nu_{\mu}$ and $\bar{\nu}_{\mu}$ data, simultaneously varying the three model parameters $m$, $M$, and $V$.  The parameters are restricted to the ranges
\begin{align}
10^{-4} \textrm{ eV } < \,\,&m < 1 \textrm{ eV } \\
0.1 \textrm{ eV } < \,\,&M < 10 \textrm{ eV } \\
0.25\cdot10^{-4} \textrm{ neV } < \,\,&V < 10 \textrm{ neV }
\end{align}
and each parameter is sampled from a logarithmic distribution (so that, for example, the same number of points are sampled for $M$ in the ranges $(0.1, 1)$ eV and $(1, 10)$ eV).  We use as our $\chi^2$ function
\be
\chi^2 = 2\left(N_{\tiny{\textrm{exp}}} - N_{\tiny{\textrm{obs}}} + N_{\tiny{\textrm{obs}}}\ln\left(\frac{N_{\tiny{\textrm{obs}}}}{N_{\tiny{\textrm{exp}}}}\right)\right)
\ee
which is more suitable for samples with small statistics \cite{Amsler:2008zzb}.  To narrow down the range of parameter space, we use an approximation the $\chi^2$ value to estimate which regions of parameter space will provide good fits to the data.  A sampling of approximately $2.5\cdot10^5$ points was sufficient to fill out the parameter space and identify the 68.3\%, 95.5\%, and 99.7\% confidence level contours.  A more detailed sampling was used to identify the best fit point.

The MINOS data is divided into 24 total bins, 17 in the neutrino data and 7 in the antineutrino data.  Fitting the three parameters $m$, $M$, and $V$ yields a best fit point
\be
m = 0.0394 \textrm{ eV, } \quad M = 0.157 \textrm{ eV, } \quad V = 2.01\cdot10^{-4} \textrm{ neV.}
\ee
The best fit point has a total $\chi^2 = 24.8 = 1.24$/dof.  Note that the value of the potential at the best fit point is about 8 times the neutral current contribution to the potential $V$, which is $0.25\cdot10^{-4}$ neV.  The mostly-active mass $m$ at the best fit point is close to the corresponding value of $(\Delta m_{\textrm{atm}}^2)^{1/2}$ in the standard picture, 0.049 eV.

\begin{figure}[htbp]
\subfloat[\large{$M$ vs. $m$}] {\label{mM} \includegraphics[width = .32\textwidth] {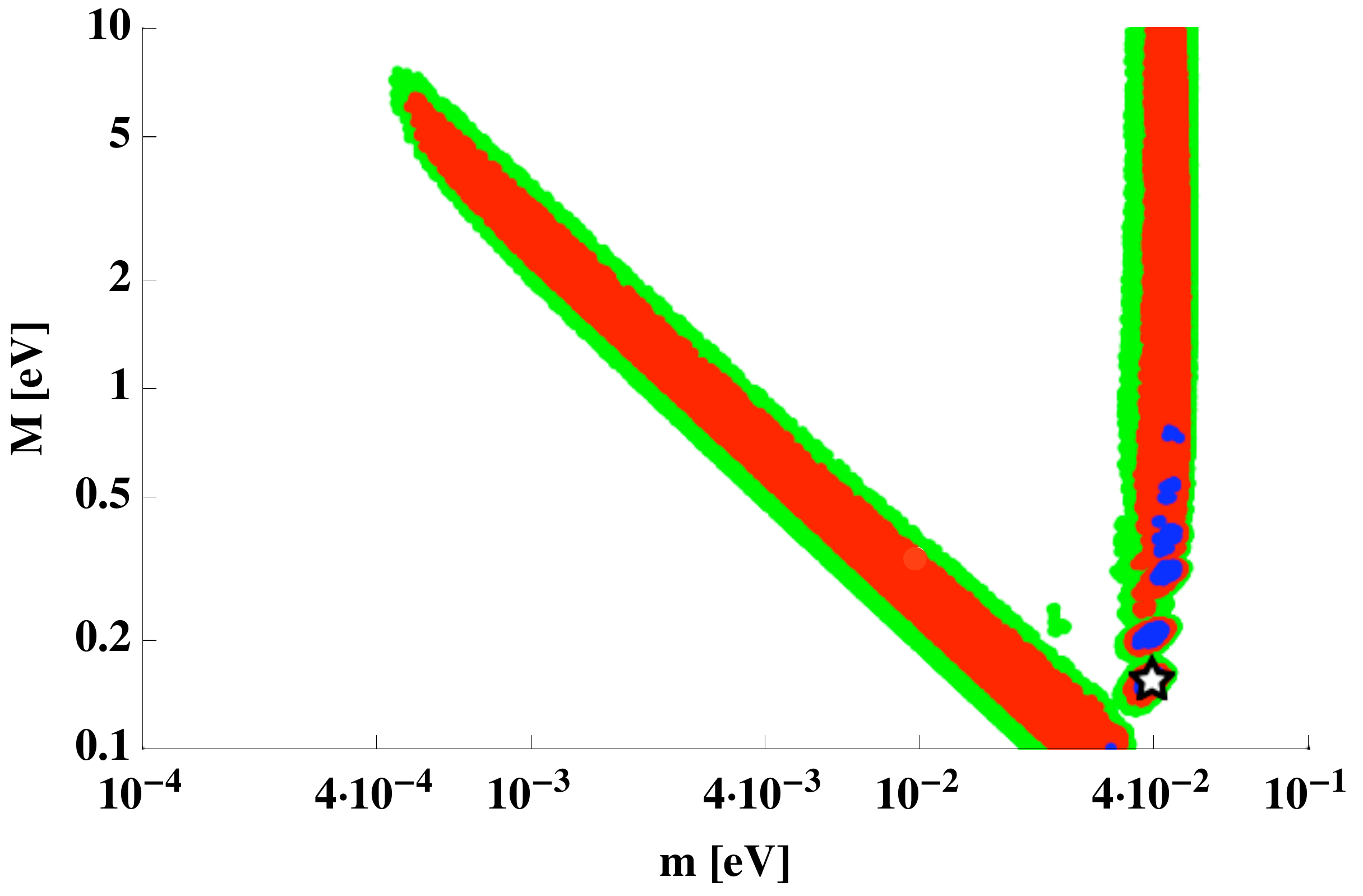}}
\subfloat[\large{$V$ vs. $m$}] {\label{mV} \includegraphics[width = .32\textwidth] {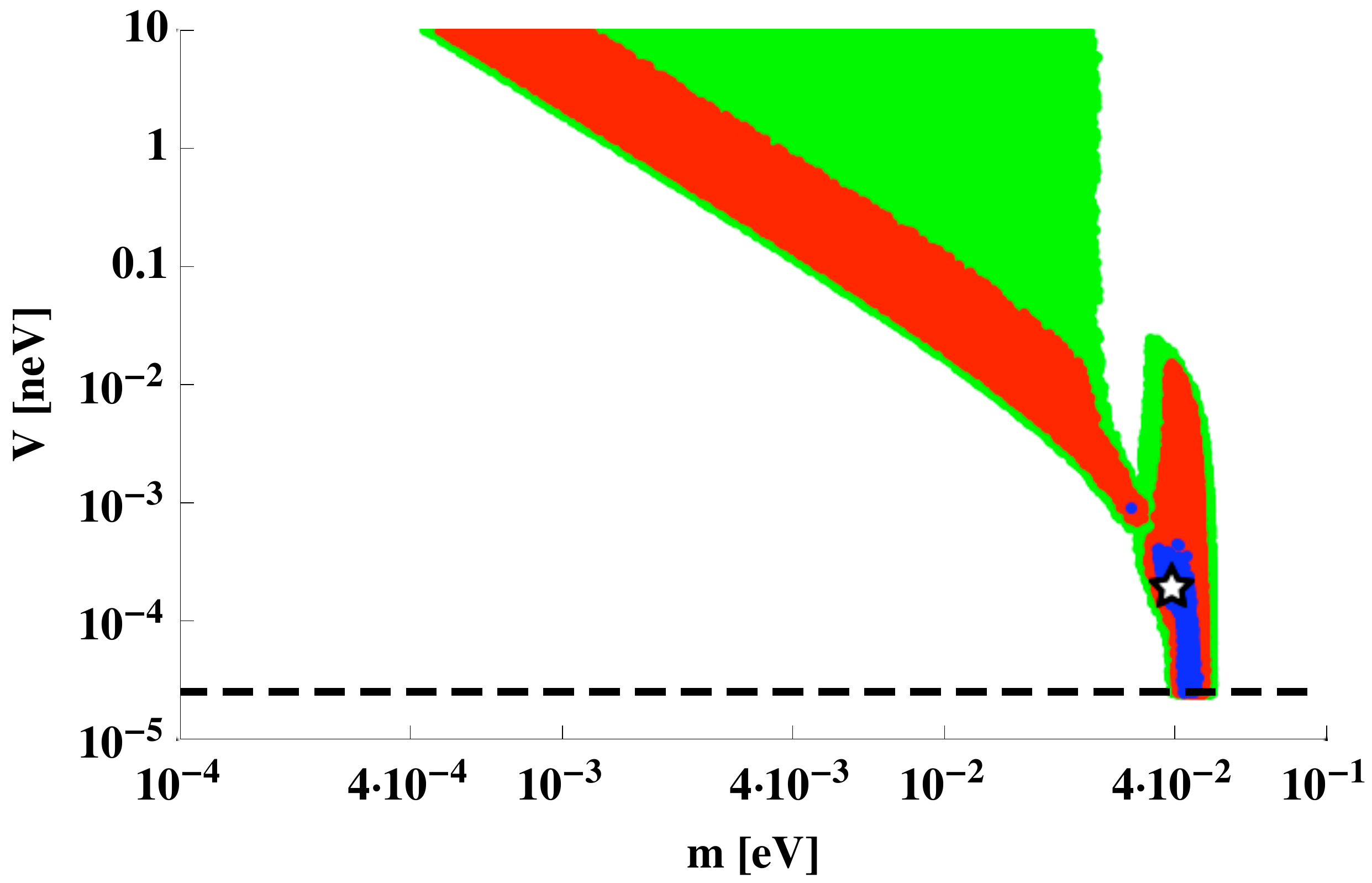}}
\subfloat[\large{$V$ vs. $M$}] {\label{MV} \includegraphics[width = .32\textwidth] {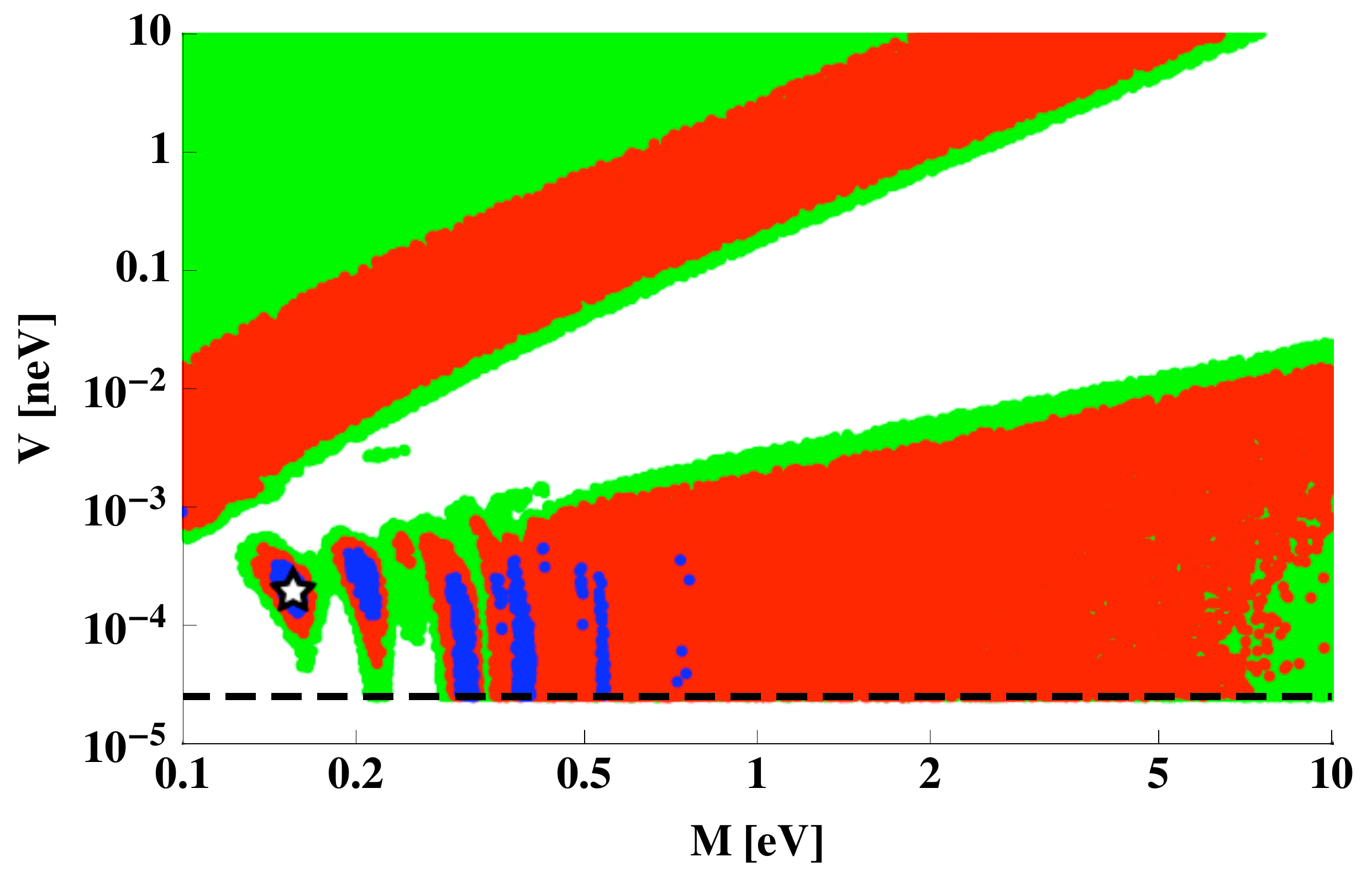}}
\caption{Regions within $68\%$ (blue), $95.5\%$ (red), and $99.7\%$ (green) confidence limits, along with the best fit point (star), for pairs of parameters.  The neutral current contribution to $V$, which is the lower limit, is the dashed line.}
\label{ParamSpace}
\end{figure}

In Fig.~\ref{ParamSpace}, we show the regions within the $68.3\%$, $95.5\%$, and $99.7\%$ confidence limits of the best fit point for pairs of parameters.  In each plot, we have projected out the other parameter, taking the minimum $\chi^2$ over the parameter not displayed.  The figures clearly indicate there are two regions of parameter space that are favored by the model.  Both regions have $\tilde{m}^2$ near $\Delta m_{\textrm{atm}}^2$ for energies near the oscillation dip for both neutrinos and anti-neutrinos.

The two regions of parameter space fitting the MINOS data have very different characteristics.  One region has a nearly constant value of $m$, approximately the best fit value of $m$, and prefers larger values of $M$ and smaller mixing with the sterile neutrino.  This region contains the natural extrapolation of the model to the standard picture, where $m^2 = \Delta m_{\textrm{atm}}^2$ and $M\to\infty$, which removes mixing between active and sterile neutrinos. 

The second favored region has larger values of $V$ and smaller values of $m$. In the second region $m$ and $M$ are inversely related: as $m$ increases in this region, $M$ decreases. The allowed parameters produce an oscillation maximum for neutrinos at the MINOS baseline at about the same energy as the standard picture, while antineutrinos have a  nonstandard looking pattern.  In the second region the oscillation pattern for neutrinos can differ significantly from the standard picture in low density material and for neutrino oscillations at different baselines. 

Two other experiments with evidence for muon neutrino disappearance are K2K and SuperK.   K2K uses a muon neutrino beam with baseline of 250 km and energy range from 0.5 GeV to 3 GeV. SuperK detects neutrinos produced by cosmic rays in the atmosphere and infers a baseline distance from their direction, with upward going neutrinos traveling much farther than downward going neutrinos.   SuperK cannot distinguish neutrino events from antineutrinos, however most of their events are from neutrinos due to the larger neutrino cross section. K2K does not have any antineutrino data.  SuperK data prefers oscillations of muon neutrinos into tau neutrinos over sterile states \cite{Fukuda:2000np}. In our model muon neutrinos at SuperK should mostly oscillate into $\tau$ neutrinos, since the mixing angle of neutrinos with sterile neutrinos is small at all energies and matter densities. However,  due to the enhanced mixing angle of antineutrinos with sterile neutrinos in matter, for some parameters a significant fraction of the  upward going muon antineutrinos at SuperK could convert  into sterile states. 

We do not attempt an analysis of constraints from the K2K data, because the statistics are very limited, or of the SuperK data, because such an analysis requires access to details of the data. However we note that 
the first  region of our fit looks very much like  the standard picture at shorter baseline experiments such as K2K, where the effects of the potential are small, and in low density material, such as experienced by downward going neutrinos at SuperK, since in the first region the potential is small, mixing with the sterile state is small,  and muon neutrino oscillations are dominated by $m$, which is near the standard atmospheric value. 
In the second region however the potential is larger and the neutrino oscillation pattern looks nonstandard at all distances  other than the MINOS baseline. This region could conceivably be constrained by an analysis of SuperK data.  For instance, the muon neutrino survival probability for downward going neutrinos at SuperK  will be higher than the standard picture because matter effects are small in the atmosphere and because the vacuum mass squared difference between the neutrinos with large mixing angle is smaller. A SuperK analysis was done for mass varying neutrinos \cite{us,Kaplan:2004dq} whose mass was different in the earth and in the atmosphere, and this analysis did produce some constraints \cite{Abe:2008zza}.  In particular, a model with no muon neutrino oscillations in air was excluded at a 3.5 $\sigma$ level. Thus a similar analysis should be able to exclude the regions of parameter space with very small $m$. 

All of our fit points are consistent with short baseline tests of muon neutrino disappearance. The strongest such constraints come from CDHS \cite{Dydak:1983zq} which is sensitive to mass squared differences larger than 0.3 eV$^2$, and has the strongest mixing angle constraints for muon neutrino mixing for a mass squared difference  of 2.5 eV$^2$. For this model the we find the CDHS constraints are weaker than those of MINOS, due to the fact that the sterile mixing angle is quite small for larger mass squared differences. For comparison, we give the   the effective mixing angle $\theta_{\rm eff} $ for 2 neutrino interpretations of muon neutrino disappearance  at short baseline:
\be
\sin^2(2\theta_{\rm eff})\approx\sin^2\theta_s(1+\cos^2\theta_s) \ .
\ee
For the case where $\alpha$ is negligible, the fit points always have  $\sin^2(2\theta_{\rm eff})$    smaller than  $0.13$ for  effective mass squared differences larger than $0.3$ eV$^2$. For effective mass squared differences larger than 2 eV$^2$, $\sin^2(2\theta_{\rm eff})< 0.05$, compared with the strongest CDHS constraint of $\sin^2(2\theta_{\rm eff})< 0.053$.  Nonnegligible $\alpha$  makes the effective neutrino mixing angle  smaller. Thus the CDHS results do  not provide any additional constraint.

We also remove the $B-L$ potential from the model, so that only the neutral current potential is present, and perform a two parameter fit to $m$ and $M$.  This fit finds a minimum $\chi^2 = 28.1 = 1.34$/dof, and lies within the $68\%$ confidence level contour of the best fit point for the three parameter fit.  The best fit values are
\be
V_{B-L} = 0:\, m = 0.0420 \textrm{ eV, } \,\, M = 0.309 \textrm{ eV.}
\ee
and  in good agreement with the MINOS neutrino and antineutrino data.

\begin{figure}[htbp]
\subfloat[\large{neutrinos}] {\label{nu} \includegraphics[width = .48\textwidth] {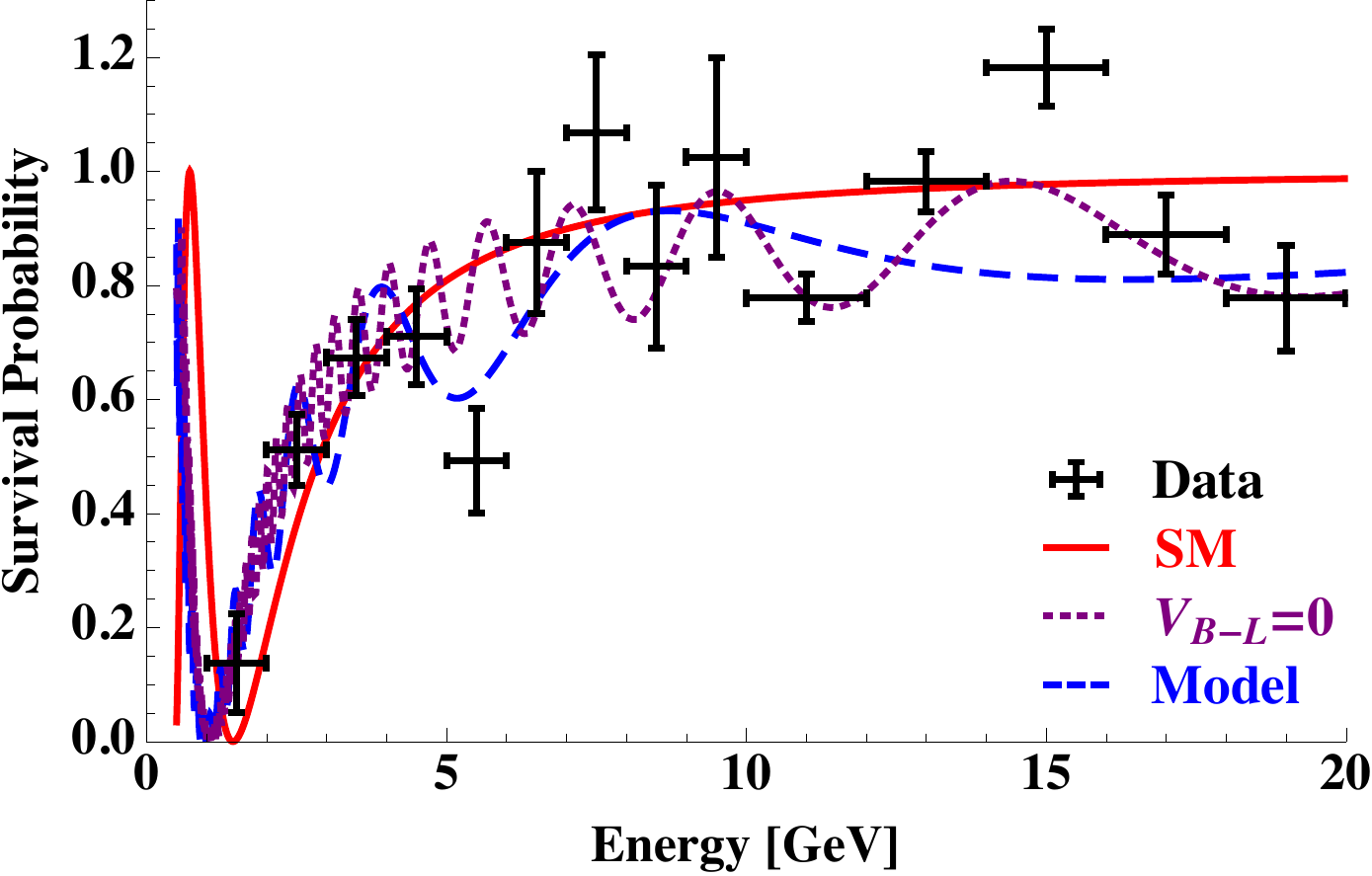}}
\subfloat[\large{antineutrinos}] {\label{nubar} \includegraphics[width = .48\textwidth] {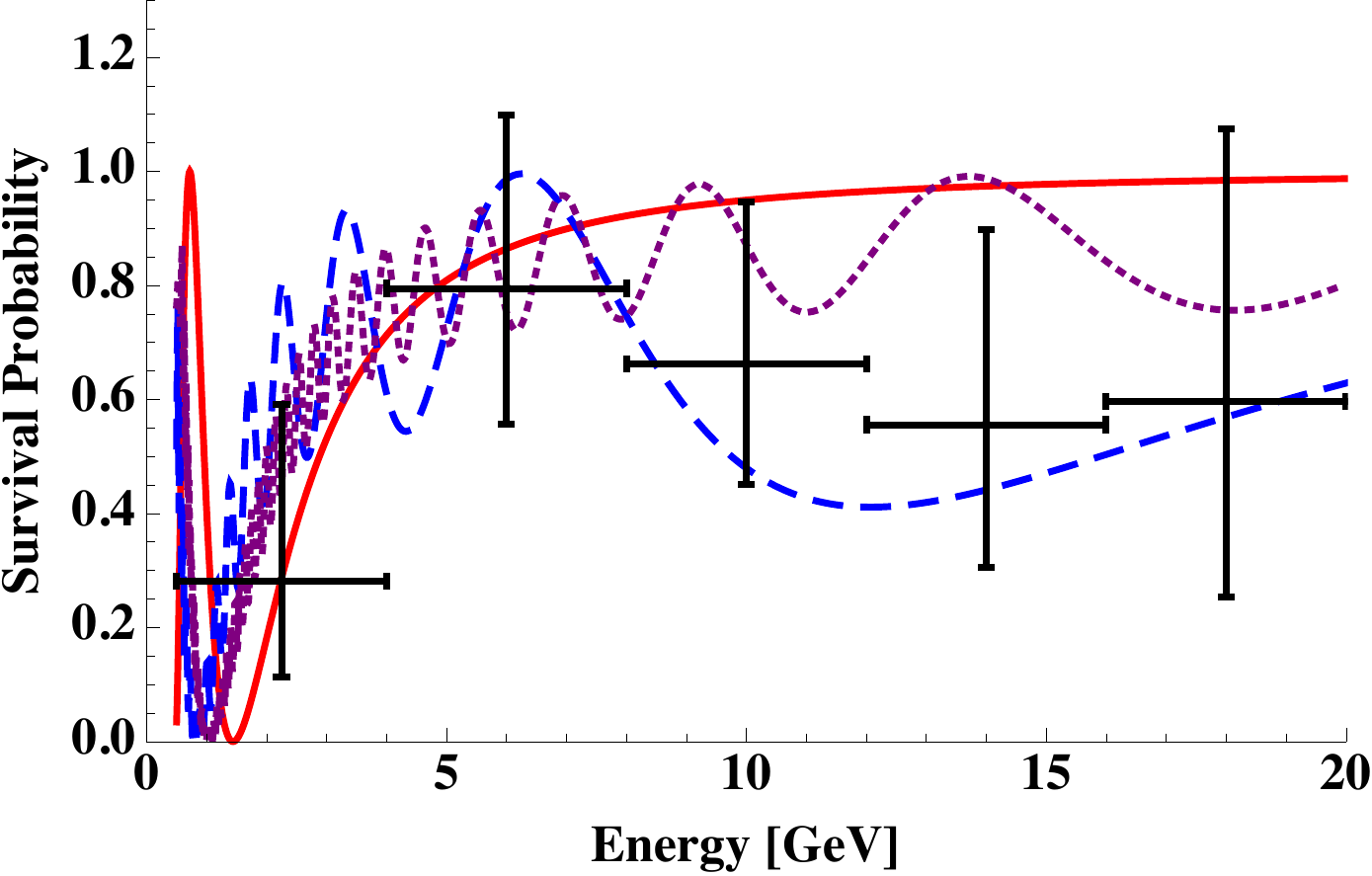}}
\caption{Muon flavor survival  probability for the various models (solid red is the standard picture, dashed blue is the model with $V_{B-L}$, and dotted purple is the model without) against the data as a function of energy.}
\label{ModelSpectra}
\end{figure}

In Fig.~\ref{ModelSpectra}, we plot the predicted muon flavor survival probabilities as a function of energy at MINOS for the model at the best fit points with and without $V_{B-L}$ against the data.  For reference, we also include the  survival probability for the standard picture.

\begin{figure}[htbp]
\subfloat[\large{neutrinos, best fit $V_{B-L}$  }] {\label{nuV} \includegraphics[width = .48\textwidth] {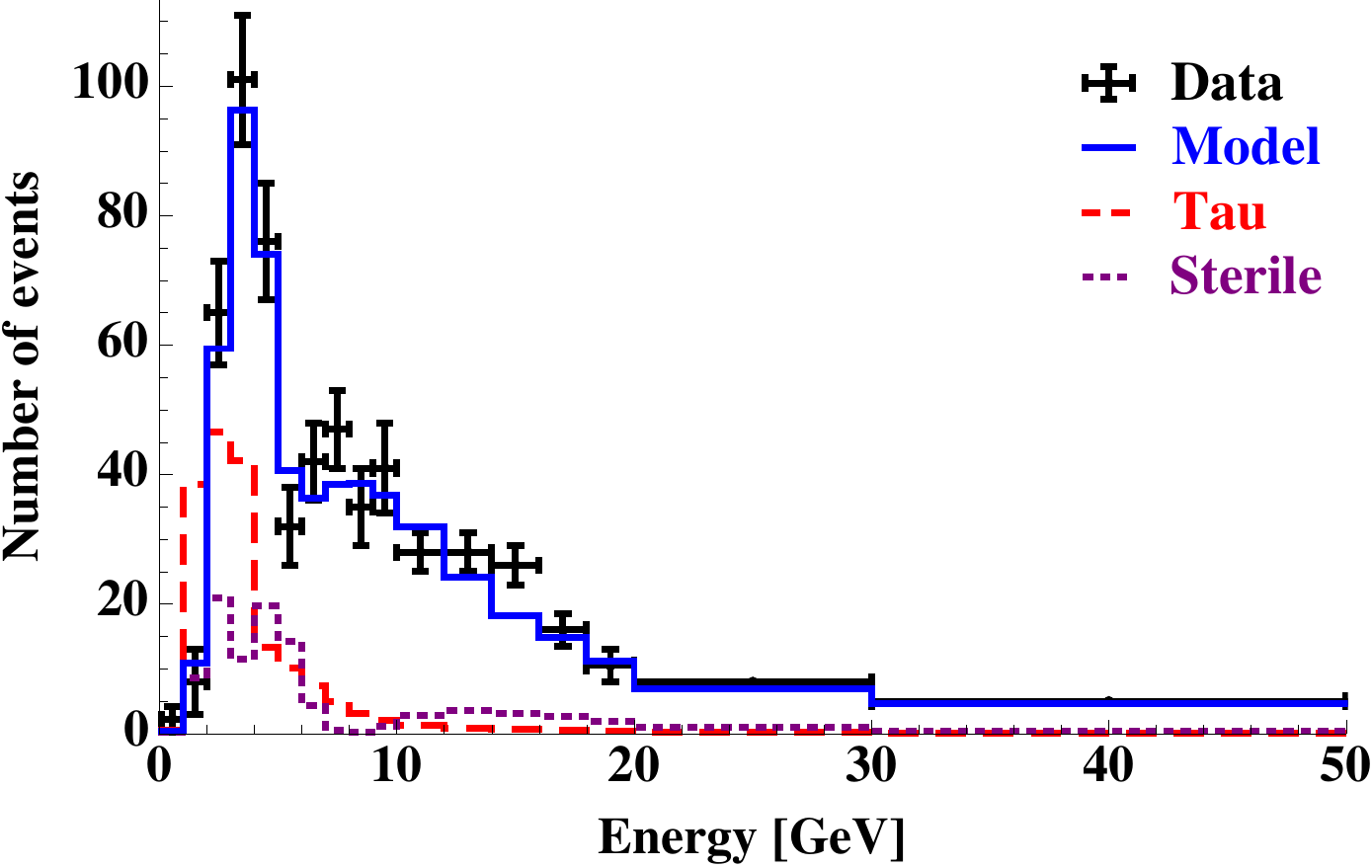}}
\subfloat[\large{antineutrinos, best fit $V_{B-L}$  }] {\label{nubarV} \includegraphics[width = .48\textwidth] {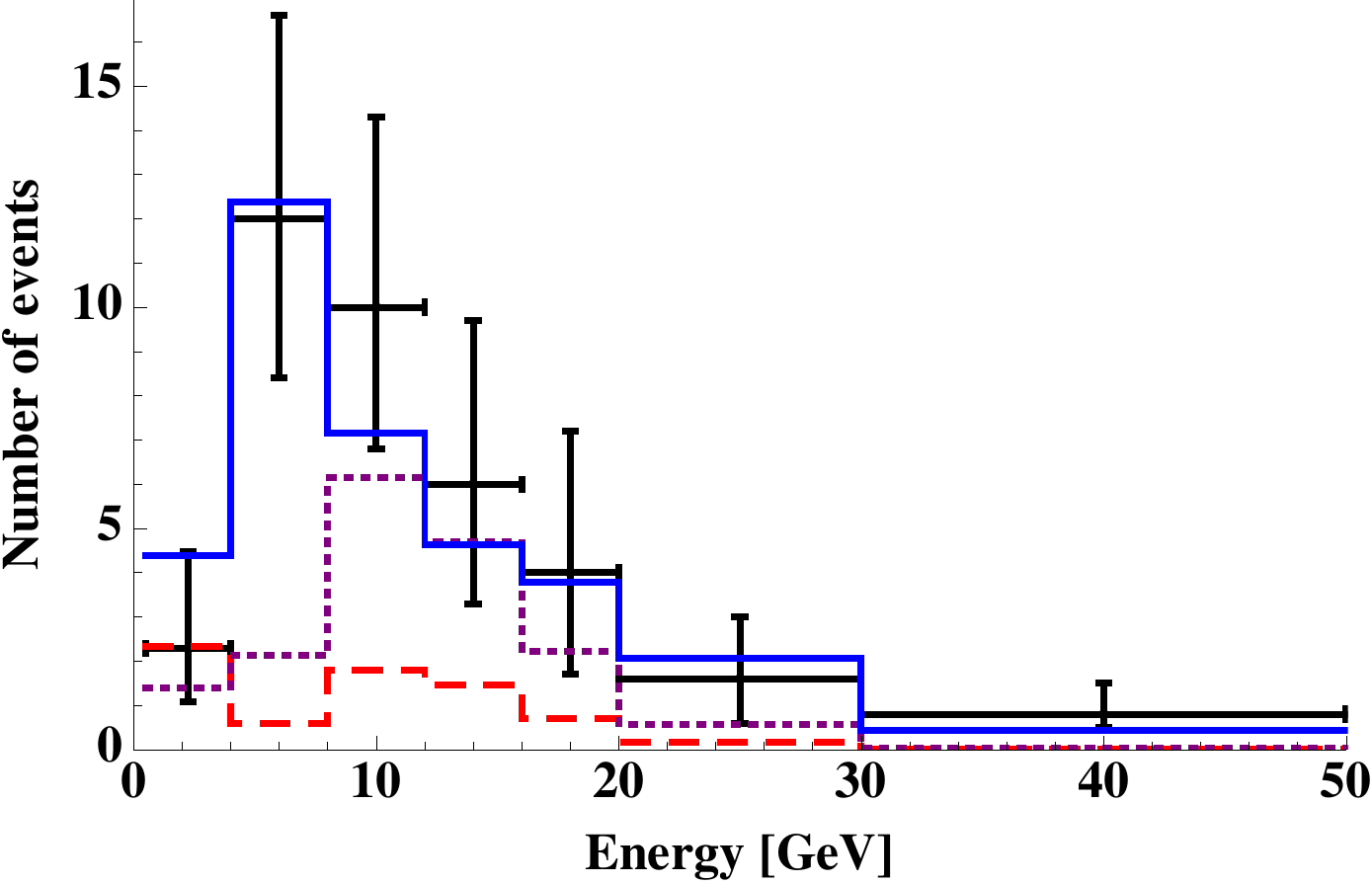}}

\subfloat[\large{neutrinos, no $V_{B-L}$}] {\label{nuNoV} \includegraphics[width = .48\textwidth] {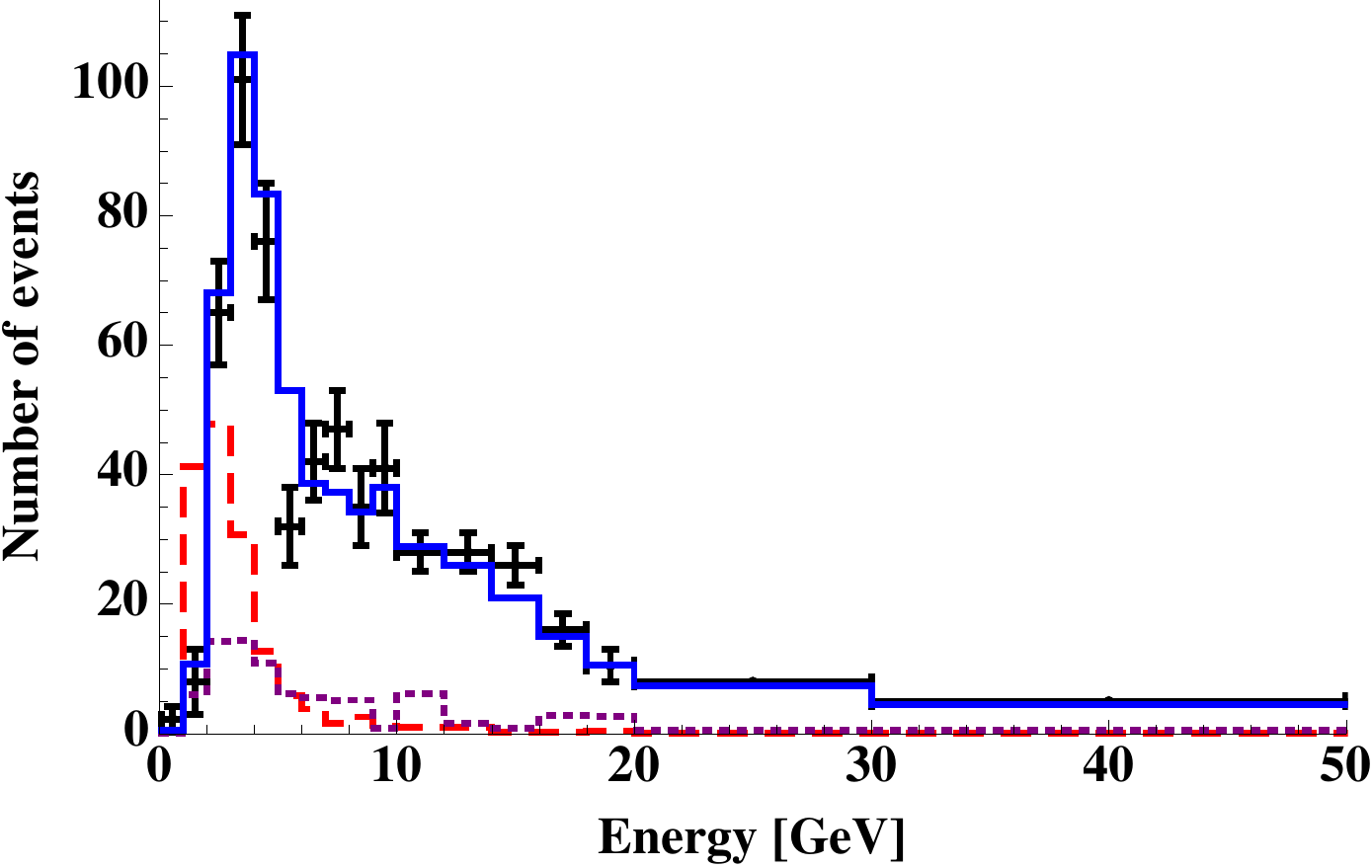}}
\subfloat[\large{antineutrinos, no $V_{B-L}$}] {\label{nubarNoV} \includegraphics[width = .48\textwidth] {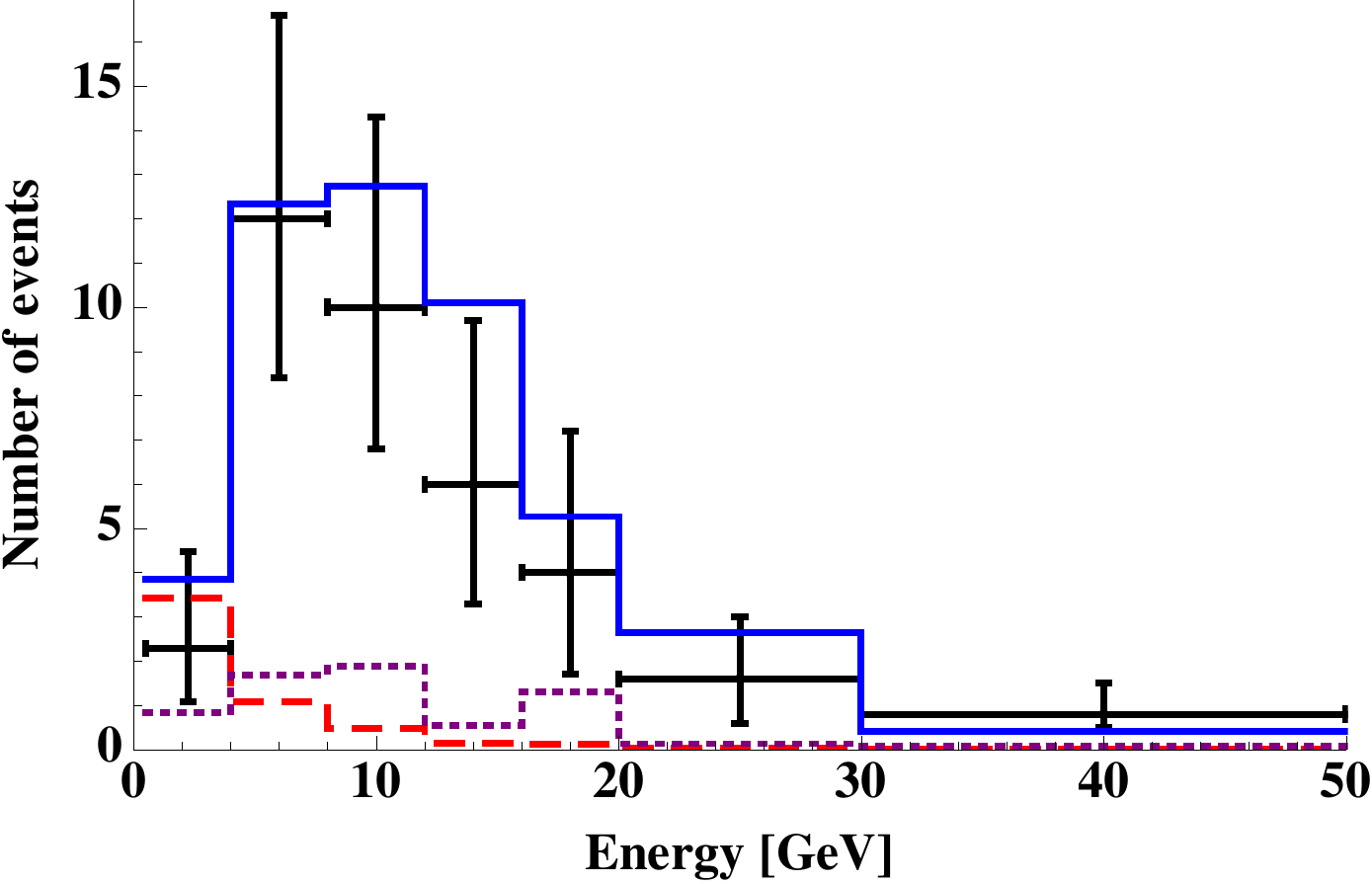}}
\caption{Predicted event rates with data as a function of neutrino energy, for the best fit point.  The solid blue curve shows the $\mu$ flavor survival rate, while the dashed red curve and dotted purple curve show the rates for conversion into $\tau$ and sterile flavors respectively.}
\label{MuTauSterile}
\end{figure}

An analysis of the MINOS neutrino data has shown that oscillation into $\nu_\tau$ is preferred over $\nu_s$.  The model is consistent with this observation, as neutrinos oscillate primarily into $\nu_\tau$. Antineutrino  have larger oscillations into the  sterile flavor than do neutrinos, and for the best fit parameters  oscillations of muon antineutrinos  into sterile neutrinos dominate oscillations in to $\bar{\nu}_\tau$ at higher energies.  In Fig.~\ref{MuTauSterile}, we plot the $\tau$ and sterile flavor conversion rates for our model at the best fit point for both neutrinos and antineutrinos at MINOS.  In the same figure we also show the same data at the best fit point of the model with  $V_{B-L}$ set to zero.
Note that the rate of conversion of active neutrinos into sterile is predicted to be small, due to the small  vacuum active-sterile mixing angle $\theta_{34}$. However the rate for active antineutrinos oscillations into sterile  neutrinos is enhanced in matter.

The MINOS experiment has also searched for oscillations into sterile neutrinos via measurements of the neutral current rate at the far detector \cite{Koskinen:2009zz,Sousa:2009wp,Adamson:2010wi}. They report a ratio $R$ of observed to expected neutral current events. The effect of  $\nu_e$ appearance, which can mimic a neutral current,  is treated as an uncertainty in the analysis. The result  is
\beq
R&=0.99\pm0 .09({\rm stat})\pm 0.07({\rm syst}) (-0.08) \ \ \ &E<3\ \gev\cr
R&=1.09\pm0.12({\rm stat})\pm0.10 ({\rm syst})(-0.13)\ \ \ &3<E<120\ \gev\cr
\eeq
where $\nu_e$ appearance is assumed to be negligible, and  the last number in parentheses shows the effect on $R$ of assuming the maximally-allowed $\nu_e$ appearance. These results are obtained from a neutrino run, with only a 7\% antineutrino component to the beam.
We have also computed $R$ for the best fit point, with and without a $B-L$ potential,  in our model. We have neglected the possibility of sterile neutrino scattering via the $B-L$ gauge boson and mimicking a neutral current, because we are assuming the $B-L$ coupling constant is small enough so that this is negligible. We find:
\beq
R&=0.88 \ \ \ &E<3,\ \gev,\  {\rm neutrinos,\  best \ fit \ }V_{B-L}\cr
R&=0.83\ \ \ &E<4,\ \gev,\  {\rm antineutrinos,\  best \ fit \ }V_{B-L}\cr
R&=0.89\ \ \ &3<E<50, \ \gev,\ {\rm neutrinos,\  best \ fit \ }V_{B-L}\cr
R&=0.71\ \ \ &4<E<50, \ \gev,\ {\rm antineutrinos,\  best \ fit \ }V_{B-L}\cr
R&=0.90 \ \ \ &E<3,\ \gev,\  {\rm neutrinos,\  no \ }V_{B-L}\cr
R&=0.90\ \ \ &E<4,\ \gev,\  {\rm antineutrinos,\  no \ }V_{B-L}\cr
R&=0.90\ \ \ &3<E<50, \ \gev,\ {\rm neutrinos,\ no\ }V_{B-L}\cr
R&=0.89\ \ \ &4<E<50, \ \gev,\ {\rm antineutrinos,\  no \ }V_{B-L}.\eeq
The antineutrino data is split up at 4 GeV energy instead of 3 because the lowest bin for the antineutrinos goes up to 4 GeV.
Note that while  the neutrino neutral current rate is in agreement with the reported MINOS data, there is an interesting possibility that the antineutrino run at MINOS could detect a significant depletion in the  neutral current rate, particularly at higher energies. 

\section{Conclusions}

There is no compelling evidence for an antineutrino oscillation anomaly at the MINOS experiment.  The data is statistics limited, and the current discrepancy is not statistically significant.  However, this discrepancy offers the opportunity to discuss models that produce interesting experimental signatures at neutrino oscillation experiments.  A neutrino/antineutrino oscillation anomaly can be explained by CPT violation, but it can also be explained by more conservative, well-motivated, and theoretically sound models.

As we have shown in this work, the addition of a sterile neutrino can generate a difference in the neutrino and antineutrino oscillation probabilities due purely to the weak neutral current interactions, and the addition of a $B-L$ gauge interaction will coherently add to the potential and enhance the oscillation probability difference.  The model we have proposed can fit well the reported MINOS neutrino and antineutrino data, and matter effects are a nice framework to explore neutrino oscillations outside of the standard picture.

As long baseline neutrino experiments enter an era of greatly increased baseline, intensity, and diversity, it is important to keep in mind that the history of neutrino experiment is full of surprises. Long baseline neutrino oscillation experiments using antineutrinos  offer a unique and sensitive window to possible exotic sectors and forces, and it is important to keep a variety of possibilities in mind when analyzing the data.

\medskip
\noindent {\bf Acknowledgments}
This work was partially supported by the DOE under contract DE-FGO3-96-ER40956. The work of Netta Engelhardt was partially supported by the NSF under the University of Washington REU program. We thank Banibrata  Mukhopadhyay, Tommy Ohlsson, and Tony Mann for correspondence.
\bibliography{neutrino}
\bibliographystyle{apsrev}

\end{document}